 \documentclass[final,5p,times,twocolumn,authoryear]{elsarticle}

\usepackage{amssymb}
\usepackage{lipsum}
\usepackage{tikz}
\usepackage{amsmath}
\usepackage{comment}
\usepackage{threeparttable}
\usepackage{float}
\usepackage{titleps}
\usepackage{rotating}  
\biboptions{sort&compress}
\usepackage{url}
\usepackage{orcidlink}
\usepackage[ruled,vlined,linesnumbered]{algorithm2e}
\usepackage{hyperref}
\hypersetup{
    colorlinks=true,
    linkcolor=blue,
    filecolor=magenta,      
    urlcolor=cyan,
    pdfpagemode=FullScreen,
    }
\usepackage{subfig}
\newcommand\I{{\rm i}}
\newcommand\red{\color{red}}

\setcitestyle{numbers,square,comma}

\newpagestyle{regular}{
  \sethead{}{}{USTC-ICTS/PCFT-24-20}
  \setfoot{}{\thepage}{}
}

\begin{document}

\pagestyle{regular}
\begin{frontmatter}



\title{A new method for finding more symmetry relations of Feynman integrals}


\author[a]{Zihao Wu\orcidlink{0000-0003-3561-5403}}
\ead{wuzihao@mail.ustc.edu.cn}
\author[b,c]{Yang Zhang\orcidlink{0000-0001-9151-8486}}
\ead{yzhphy@ustc.edu.cn}
\affiliation[a]{organization={School of Fundamental Physics and Mathematical Sciences, Hangzhou Institute for Advanced Study, UCAS},
            addressline={}, 
            city={Hangzhou},
            postcode={310024}, 
            state={},
            country={China}}
\affiliation[b]{organization={Department of Physics, University of Science and Technology of China},
            addressline={}, 
            city={Hefei},
            postcode={230026}, 
            state={},
            country={China}}
\affiliation[c]{organization={Peng Huanwu Center for Fundamental Theory},
            addressline={}, 
            city={Hefei},
            postcode={230026}, 
            state={},
            country={China}}

\begin{abstract}
We introduce a new method for deriving Feynman integral symmetry relations. By solving the ansatz of momentum transformation in the field of rational functions rather than constants, this method can sometimes find more symmetry relations, compared to some state-of-the-art software. The new method may help to further decrease the number of unique sectors in an integral family. Well-chosen gauge conditions are implemented in this method for the efficient symmetry searching.
\end{abstract}







\end{frontmatter}

\section{Introduction}
The evaluation of Feynman integrals is one of the most important topics in perturbative quantum field theory. Especially for high-precision computation of scattering amplitudes, we usually need to evaluate a large number of multi-loop Feynman integrals. This kind of computation is often of high complexity. In past several decades, Feynman integral techniques have been well developed, including different representations of Feynman integrals \cite{Smirnov:2012gma,peskin1995introduction,Lee:2013hzt,Lee:2013mka,Lee:2014tja,Baikov:1996cd,Baikov:1996rk,Baikov:2005nv,Smirnov:1999gc, Tausk:1999vh}, Feynman integral reduction \cite{Passarino:1978jh,Ossola:2006us,Ossola:2007ax,Tkachov:1981wb,
Chetyrkin:1981qh,
Laporta:2000dsw,
Gluza:2010ws,Schabinger:2011dz,Ita:2015tya,Larsen:2015ped,Bohm:2017qme,vonManteuffel:2020vjv,Bohm:2018bdy,Bendle:2019csk,Wu:2023upw,
Mastrolia:2018uzb,Frellesvig:2019uqt,Frellesvig:2019kgj,Frellesvig:2020qot,
Smirnov:2006wh,Smirnov:2006vt,Smirnov:2006tz,Smirnov:2005ky,Bitoun:2017nre,Lee:2014tja,Baikov:1996iu,Lee:2008tj,Grozin:2011mt,Smirnov_2006,Smirnov:2012gma,vonManteuffel:2014ixa,Peraro:2016wsq,Klappert:2019emp,Klappert:2020aqs,Klappert:2020nbg,Peraro:2019svx,
Kosower:2018obg,
Anastasiou:2004vj,
Smirnov:2008iw,Smirnov:2013dia,Smirnov:2014hma,Smirnov:2019qkx,
Maierhofer:2017gsa,Maierhofer:2018gpa,Maierhofer:2019goc,
Lee:2013mka,
Studerus:2009ye,vonManteuffel:2012np,Guan:2019bcx,Liu:2021wks,Guan:2024byi,Lange:2024mmz,Usovitsch:2022wvr}, differential equation methods \cite{Henn:2013pwa,Henn:2014qga,
Remiddi:1999ew,Goncharov:1998kja,Goncharov:2010jf,Schabinger:2018dyi,Dlapa:2021qsl,He:2022ctv,Henn:2021cyv,Argeri:2014qva,Wasser:2022kwg,Wasser:2018qvj,Chicherin:2018old,Henn:2020lye,Meyer:2016slj,Meyer:2017joq,Prausa:2017ltv,Gituliar:2017vzm,Dlapa:2020cwj,Lee:2014ioa,Lee:2017oca,Lee:2020zfb,Frellesvig:2017aai,Harley:2017qut}, auxiliary mass flow method \cite{Liu:2022mfb,Liu:2022chg,Liu:2021wks,Liu:2020kpc,Liu:2018dmc,Liu:2017jxz}, expansion-by-regions method \cite{Beneke:1997zp,Smirnov:2012gma}, dimensional recursion methods \cite{Lee:2017yex,Lee:2015eva,Lee:2009dh,Tarasov:1996br,Lee:2017ftw}, etc. In the above methods, Feynman integral reduction is very important. Using a set of relations between the different integrals, the reduction process decreases the number of integrals that are to be computed concerning a given problem. Wherein, the so-called integration-by-parts (IBP) reduction \cite{Tkachov:1981wb,Chetyrkin:1981qh} is one of the most critical methods, serving as a standard procedure in nowadays' work flow of Feynman integral computation. It uses integration by parts identities by considering the vanishing boundary in the integration, to reduce the given scalar integrals. After the reduction, the integrals can be expressed as a linear combination of a much smaller set of integrals, called master integrals. 

The symmetry relations between Feynman integrals are another type of important relations. They reflect the graphical symmetry of Feynman diagrams or algebraic symmetry for Feynman integral representations. These relations are very useful in the IBP reduction processes and other analytical derivations (see ref. \cite{deLima:2024hnk} for a latest research example). In the IBP reduction process, in most cases, the symmetry relations provide additional relations in an integral family other than IBP relations. Thus, they are very useful in a  reduction process, to further decrease the number of master integrals. 

In addition, symmetry relations are usually simpler than IBP relations. Thus, making use of the symmetry relations generally makes the IBP reduction more efficient. An important usage is to decrease the number of sectors that are to be concerned in an IBP reduction, where sectors are sets of integrals corresponding to different top/sub topologies of the diagrams. To achieve this, we use the \textit{symmetries between sectors}, which are symmetry relations that map all integrals in a sector to other sectors. Thus, the mapped sectors are not to be concerned in an IBP reduction. This reduces the computational cost.

The symmetry relations between sectors can be found by applying a suitable linear transformation on the internal and external momenta for Feynman integrals. 
To derive such a transformation, one needs to solve a set of equations for the coefficients of the linear transformation. The equations, mostly quadratic, are not always easy to solve. If the transformation allows rational functions in coefficients, 
there may be large degrees of freedom for the solutions. So, it is very computational expensive to find the general solution or a special solution. As a compromise, many state-of-the-art Feynman integral reduction programs, consider the momentum transformation with only constant coefficients during the symmetry relation searching. This approach may miss certain relations.

In this paper, we introduce a novel method to find symmetry relations between Feynman integrals from the momentum transformation with {\it rational function} coefficients. By introducing sophisticated solution selection conditions (called ``gauge" condition in this paper), the new method finds the momentum transformation with rational function coefficients efficiently. Using this method, we can sometimes find more symmetry relations between different sectors, compared to the state-of-the-art Feynman integral reduction software. The method introduced in this paper has been implemented in the package {\sc NeatIBP} \cite{Wu:2023upw}, and is planned to be implemented in the future version of {\sc Kira} \cite{Maierhofer:2017gsa,Maierhofer:2018gpa,Maierhofer:2019goc,Klappert:2020nbg}.

This paper is organized as follows. In Section \ref{sec:2} we include some basic concept of Feynman integral symmetry. We also give a discussion of the basic principle of the new method, and provide a typical example. In Section \ref{sec:3}, we state the first part of the method, which is deriving momentum transformation in case that the external lines are in groups. In Section \ref{sec:4}, we state the second part of the method about how to determine external momenta groups. 
In Section \ref{sec:example}, we present an example.




\section{Basic concepts}\label{sec:2}
\subsection{Feynman integral families and sectors}
The scalar integrals for a given Feynman diagram can be expressed as follows,
\begin{equation}\label{eq:FI}  I_{\alpha_1,\cdots,\alpha_n}=\int \frac{{\rm d}^D l_1}{\I \pi^{D/2}}\cdots\frac{{\rm d}^D l_L}{\I \pi^{D/2}}\frac{1}{D_1^{\alpha_1}\cdots D_n^{\alpha_n}},
\end{equation}
with $D_i$ the (inverse) propagators and $\alpha_i$ integers. The set of all integrals $I_{\alpha_1,\cdots,\alpha_n}$, with $\alpha_i\in \mathbb Z$, $i=1,\ldots, n$ is called an integral family. 

An integral family has many sectors, which reflects the $D_i$'s in the denominator. One way to label a sector is via the set of positive indices, 
\begin{equation}
    S(I_{\alpha_1,\cdots,\alpha_n}):=\{i|\alpha_i>0\}.
\end{equation}

In this notation, we call $S_a$ a sub sector of $S_b$ if $S_a\subset S_b$. 
Equivalently,  $S_b$ is a super sector of $S_a$.
\subsection{Feynman integral symmetries}
In this paper,  we mainly discuss the symmetries between two sectors of an integral family. The same idea can be used to find symmetry between two sectors of two {\it different} integral families.

Our method finds a linear transformation on the internal momenta $l_i$ and the external momenta $p_k$. This transformation maps any integral in the {\it source} sector, $S_a$, to a linear combination of a {\it image} sector $S_b$ and its sub sectors.  

The linear transformation has the following ansatz,
\begin{align}
\label{eq:loop momentum symmetry trans.}
    &l_i\mapsto l_i^\prime =\mathcal A_{ij} l_j +\mathcal B_{ik}p_k,  \\
    &p_i\mapsto p_i^\prime=\mathcal C_{ij} p_j.\label{eq:external momentum symmetry trans.}
\end{align}
where 
\begin{equation}\label{eq: momentum symmetry trans Det. condition}
     \det(\mathcal A_{ij})=\pm 1.
\end{equation}
because of the restriction on the Jacobian.

Under such transformation, we also require the scalar products of the external momenta  remain unchanged, 
\begin{equation}
    p_i\cdot p_j \mapsto p_i^\prime\cdot p_j^\prime= p_i \cdot p_j, 
    \label{eq:kinematic_constraint}
\end{equation}
and the denominator are permuted as\footnote{For the $D_i$'s in the numerator, i.e. $i\notin S_a$, we  have no requirement.}, 
\begin{equation}\label{eq:momentum symmetry trans prop. condition}
    D_i\mapsto D_i^\prime= D_{\sigma(i)}, \quad\text{for }i\in S_a,
\end{equation}
where $\sigma$ is an injective index map satisfying 
\begin{equation}
    \{\sigma(i)|i\in S_a\}=S_b,
\end{equation}
If the map $\sigma$ is known, the equations \eqref{eq: momentum symmetry trans Det. condition}  \eqref{eq:kinematic_constraint} and \eqref{eq:momentum symmetry trans prop. condition} are in general a set of nonlinear equations of undetermined coefficients $\mathcal A$, $\mathcal B$ and $\mathcal C$. 

The clue is that $\sigma$ must map the numerator-one integral of sector $S_a$ to the numerator-one integral of $S_b$. Here, a  numerator-one integral for $S_a$ refers to the integral,
\begin{equation}\label{eq:corner_integral}  \int \frac{{\rm d}^D l_1}{\I \pi^{D/2}}\cdots\frac{{\rm d}^D l_L}{\I \pi^{D/2}}\frac{1}{D_1^{\beta_1}\cdots D_n^{\beta_n}},
\end{equation}
with the $\beta_i\geq1$ if $i\in S_a$, and otherwise $\beta_i=0$. In other words, $\sigma$ maps the Lee-Pomeransky polynomial \cite{Lee:2013mka,
Lee:2013hzt,
Lee:2014ioa} of the sector $S_a$ to that of the sector $S_b$. So, the two Lee-Pomeransky polynomials differ only by a variable re-definition. 

Polynomials related by a variable redefinition can be identified via the Pak algorithm \cite{Pak:2011xt}. Therefore, an index map $\sigma$ can be found by running Pak algorithm over Lee-Pomeransky polynomials. 

However, our goal is to find the symmetry map between two sectors, not just between their numerator-one integrals. The information about the index map $\sigma$ is not enough. In principle, given an index map $\sigma$, we need to solve equations \eqref{eq: momentum symmetry trans Det. condition}, \eqref{eq:kinematic_constraint} and \eqref{eq:momentum symmetry trans prop. condition}  to see if such a symmetry map exists. 




\subsection{Solving the ansatz equations}
As we have stated above, given a permutation $\sigma$, we need to solve the corresponding ansatz equations \eqref{eq: momentum symmetry trans Det. condition}, \eqref{eq:kinematic_constraint} and \eqref{eq:momentum symmetry trans prop. condition} where the entries of matrices $\mathcal{A}$, $\mathcal{B}$ and $\mathcal{C}$ are unknowns. 
In some state-of-the-art reduction software, the unknowns are treated as constants. However, since the equations usually contain parameters like Mandelstam variables and masses, the solutions should in general also be functions of these parameters. Treating $\mathcal A$, $\mathcal B$ and $\mathcal C$ as constants may miss some solution.

In the method we present in this paper, the unknowns are treated as functions.  Sometimes, doing so will make the equations difficult to solve, partly because the solution may have the continuous degree of freedoms. The new method will introduce some well-chosen gauge conditions to fix them, and produce a nice solution efficiently. 



\begin{figure}[hbtp]
\centering

\begin{tikzpicture}[scale=0.8]
\draw (0,0) arc (120:60:2);
\draw(0,0) arc (-120:-60:2);
\draw(2,0)--(0,0);
\draw(2,0)--(3,0.5);
\draw(2,0)--(3,0);
\draw(2,0)--(3,-0.5);

\draw(-1,-0.5)--(0,0);
\draw(-1,0.5)--(0,0);
\draw(-1,0)--(0,0);

\node[font=\large\bfseries] at (-1.5,-1){$1$};
\node[font=\large\bfseries] at (-1.5,1){$3$};
\node[font=\large\bfseries] at (-1.5,0){$2$};
\node[font=\large\bfseries] at (3.5,0){$5$};
\node[font=\large\bfseries] at (3.5,1){$4$};
\node[font=\large\bfseries] at (3.5,-1){$6$};
\end{tikzpicture}

\caption{An apparent six-point, but in fact two-point diagram}
\label{fig:2}
\end{figure}
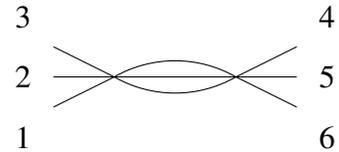

The cases for solutions with continuous degrees of freedom usually appear for low sub sectors in a family, i.e., integrals with fewer propagators.
The external momenta may form some groups, connecting to the same vertex of the diagram. So, it is in fact a diagram with fewer external lines (see Fig. \ref{fig:2}). In Section \ref{subsec:example 2}, we will present a typical example, and show that the corresponding symmetry relations are coming from linear transforms with rational function coefficients.


\subsection{An example of symmetry relation with rational function coefficients}\label{subsec:example 2}
In this sub section, we will give an example with external momenta grouped. 
Consider the diagrams in Fig. \ref{fig:1}. The independent external momenta of the two diagrams are $p_1$, $p_2$, and $p_3$. The four-point diagrams are in fact three-point diagrams, with two independent external momentum groups
\begin{equation}
    u_1=p_1+p_2, \quad u_2=p_3.
\end{equation}

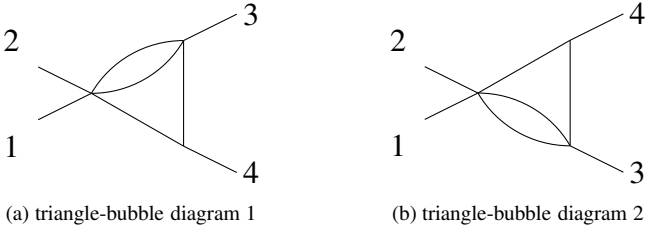
\begin{figure}[hbtp]
\centering
\subfloat[triangle-bubble diagram 1]{
\begin{tikzpicture}[scale=0.7]
\draw (0,0) arc (150:90:2);
\draw (0,0) arc (-90:-30:2);

\draw(1.732,-1)--(0,0);

\draw(1.732,-1)--(1.732,1);
\draw(1.732,1)--(2.732,1.5);
\draw(1.732,-1)--(2.732,-1.5);

\draw(-1,-0.5)--(0,0);
\draw(-1,0.5)--(0,0);

\node[font=\large\bfseries] at (-1.5,-1){$1$};
\node[font=\large\bfseries] at (-1.5,1){$2$};
\node[font=\large\bfseries] at (3,1.5){$3$};
\node[font=\large\bfseries] at (3,-1.5){$4$};

\end{tikzpicture}
}
\hspace{35pt}
\subfloat[triangle-bubble diagram 2]{
\begin{tikzpicture}[scale=0.7]
\draw (0,0) arc (-150:-90:2);
\draw (0,0) arc (90:30:2);

\draw(1.732,1)--(0,0);

\draw(1.732,-1)--(1.732,1);
\draw(1.732,1)--(2.732,1.5);
\draw(1.732,-1)--(2.732,-1.5);

\draw(-1,-0.5)--(0,0);
\draw(-1,0.5)--(0,0);

\node[font=\large\bfseries] at (-1.5,-1){$1$};
\node[font=\large\bfseries] at (-1.5,1){$2$};
\node[font=\large\bfseries] at (3,1.5){$4$};
\node[font=\large\bfseries] at (3,-1.5){$3$};

\end{tikzpicture}
}
\caption{An example of external momentum grouping}

\end{figure}\label{fig:1}

Either by looking at the diagram or by solving the determinant and propagator constrain equations, which are \eqref{eq: momentum symmetry trans Det. condition} and \eqref{eq:momentum symmetry trans prop. condition}, we conclude that the symmetry transformation for the external momentum groups is
\begin{equation}\label{eq: example 1: group transform}
    u_1 \mapsto u_1,\quad u_2\mapsto -u_1-u_2,
\end{equation}
After that, we need to find a transformation for $p_i$, i.e. \eqref{eq:external momentum symmetry trans.}, that satisfies \eqref{eq: example 1: group transform} and the external momenta constrain \eqref{eq:kinematic_constraint}. For this example, the corresponding equations have no solutions with constant coefficients. But, a solution with rational functions can be found,
\begin{equation}
\begin{aligned}
    &p_1\mapsto p_1-\frac{s+2t}{s}({p_1+p_2+2p_3}),\quad\\
    &p_2\mapsto p_2+\frac{s+2t}{s}({p_1+p_2+2p_3}),\quad \\
    &p_3\mapsto -p_1-p_2-p_3
\end{aligned}
\end{equation}
In the next sections, we introduce the method for finding such kind of solutions, by avoiding the continuous degree of freedoms.

\section{The external momentum transformation in groups}\label{sec:3}

The work flow of our method is as follows. Firstly, we determine the external momentum groups $u_i$'s\footnote{The algorithm of doing so is introduced in Section \ref{sec:4}}. After that, solve the determinant and propagator constrain equations \eqref{eq: momentum symmetry trans Det. condition} and \eqref{eq:momentum symmetry trans prop. condition},  with the  external momenta $p_i$'s replaced by the momentum groups $u_i$'s. The solutions of these new equations, according to our experience, are without continuous degree of freedom. Using computational algebraic geometry tools such as the Groebner basis \cite{CLO}, the equations without continuous degrees of freedom can be solved efficiently. The solution of the transformation reads,
\begin{align}
\label{eq:loop momentum symmetry trans. with momentum group}
    &l_i\mapsto l_i^\prime=\mathcal A^\prime_{ij} l_j +\mathcal B^\prime_{ik}u_k,  \\
    &u_i\mapsto u^\prime_i=\mathcal  C^\prime_{ij} u_j,\label{eq:external momentum symmetry trans. with momentum group}
\end{align}
where the unknowns $\mathcal{A}^\prime$, $\mathcal{B}^\prime$ and $\mathcal{C}^\prime$ are determined. After that, we try to find a transformation of the original external momenta  
that satisfy \eqref{eq:external momentum symmetry trans. with momentum group} and \eqref{eq:kinematic_constraint}. In this step, we encounter the continuous degree of freedom. Thus, we will introduce well-chosen gauges, to give a nice particular solution. In this section, we introduce two gauge choices (both implemented in {\sc NeatIBP} \cite{Wu:2023upw}), the so-called {\it delta plane projection gauge}, and the {\it orthogonalization gauge}. 

\subsection{The delta plane projection gauge}\label{subsec:dpp}
Let $u_i$ to be the grouped external momenta before the transformation and $u_i^\prime$ to be the transformed ones. Define the difference of each group as 
\begin{equation}\label{eq:delta ui}
    \Delta u_i=u_i^\prime-u_i.
\end{equation}
We can choose the linearly independent ones of the difference to form a basis $\omega_i$. Without the loss of generality we label them as the first to the $n$-th ones,
\begin{equation}
    \omega_i=\Delta u_i,\quad i\leq n,
\end{equation}
and the rest are 
\begin{equation}
    \Delta u_a={C_a}^{j}\omega_j,\quad a> n,j\leq n.
\end{equation}

In the following paragraphs, we call the linear transformation found under the delta-plane projection gauge as $T_{\text{dpp}}$. For any $k$ as a linear combination of $p_i$'s,   $k$ transforms to $T_{\text{dpp}}(k)$. 
$T_{\text{dpp}}$ is required to satisfy the following condition,
\begin{equation}\label{eq: delta plane projection additional gauge}
k=T_{\text{dpp}}(k)\Leftrightarrow k\cdot \omega_i=0, \, \forall i.
\end{equation}
For any momenta like $k$, we consider the projection to the linear space spanned by $\omega_i$'s and the  orthogonal component, i.e.
\begin{equation}
    k=k^\parallel+k^\perp.
\end{equation}
and explicitly,
\begin{equation}
    k^\parallel=\sum_{i=1}^n(k\cdot\omega^i)\omega_i,
\end{equation}
where
\begin{equation}\label{eq: w^i def}
    \omega^i=\sum_{i=1}^n(G^{-1})^{ij}\omega_j,
\end{equation}
and the matrix $G^{-1}$ is the inverse local gram matrix such that
\begin{equation}\label{eq:G^-1 def}
   \sum_{j=1}^n(\omega_i\cdot\omega_j) (G^{-1})^{jk}={\delta_i}^k.
\end{equation}
In the following discussions, we will use the local gram matrix to raise and lower the indices. With the above definitions, the following properties follow,
\begin{itemize}
    \item $k^\perp\cdot \omega_i=0,\, \forall i.$
    \item $\omega_i^\perp=0$ for $i\leq n$. Thus, $(\Delta u_i)^\perp=0$, for $i\leq n $ and $i>n$.
    \item ${u_i^\parallel}$'s form a linear basis (for $i\leq n$). To prove this, assume $\exists r_i$ such that $\sum_{i=1}^n r_i{u_i^\parallel}=0$. Considering that $u_r=\sum_{i=1}^n r_i{u_i}$, we have $u_r^\parallel=0$. Thus, $u_r^\perp\cdot \omega_i=0$. Since we have \eqref{eq: delta plane projection additional gauge}, we have $0=\Delta u_r=\sum_{i=1}^n r_i{\omega_i}$. Since $\omega_i$'s form a linear basis, we have $r_i=0$. 
\end{itemize}
We then expand $u_i$ to the $\omega_i$ basis as
\begin{equation}\label{eq:ui para to wi}
    u_i^\parallel=\sum_{j=1}^n {A_i}^j \omega_j,
\end{equation}
where ${A_i}^j=u_i\cdot \omega^j$. Because $\{u_i^\parallel \}$ is a linearly independent basis, $A$ is invertible. We can also expand $u_i^\prime$ to the basis as,
\begin{equation}\label{eq:uip para to wi}
    u_i^{\prime\parallel}=\sum_{j=1}^n  {A_i^\prime}^j\omega_j,
\end{equation}
where\footnote{Thus the matrices $A$ and $A^\prime$ commute.}
\begin{equation}\label{eq Apij def}
    {A_i^\prime}^j ={A_i}^j+{\delta_i}^j\,,
\end{equation}
Therefore,
\begin{equation}
     u_i^{\prime\parallel}=\sum_{j=1}^n{B_i}^j u_j^{\parallel},
\end{equation}
where
\begin{equation}\label{eq:Bij def}
    {B_i}^j=\sum_{k=1}^n{A_i^\prime}^k{(A^{-1})_k}^j=\sum_{k=1}^n{(A^{-1})_i}^k{A_k^\prime}^j
\end{equation}
We claim that the following transform is a solution for \eqref{eq:kinematic_constraint} and \eqref{eq: delta plane projection additional gauge}, which is, for a momenta $k$, it transforms (linearly) as
\begin{equation}\label{eq:delta plane projection transformation}
    k\mapsto T_{\text {dpp}}(k):=k^\perp+\sum_{i,j=1}^n (k\cdot \omega^i) {B_i}^j \omega_j.
\end{equation}
The proof of this claim is given in \ref{subsec: appendix dpp proof}. 

\subsection{The orthogonalization gauge}\label{subsec:ortho}
Suppose we chose linearly independent $u_i$'s, without loss of generality.  $i\leq n$, together with  orthogonal  vectors to form a basis, 
\begin{equation}
    b_i=u_i, \quad i\leq n,
\end{equation}
\begin{equation}
    b_i\cdot u_j=0, \quad i>n,j\leq n,
\end{equation}
and
\begin{equation}
    b_i\cdot b_j=\delta_{ij}, \quad i,j>n.
\end{equation}
 Similarly, we  define an orthogonal complement basis $b_i^\prime$ for $u_i^\prime$.

A vector $k$ is  expanded over the basis $\{b_i\}$ as
\begin{equation}
    k=\sum_i c_i b_i.
\end{equation}
The new gauge, which called orthorgonalization gauge, is to require the image of $k$,  $T_{\text{ortho}}(k)$, has the same coefficient with the respect to the basis $\{{b^\prime_i}\}$, i.e.
\begin{equation}\label{eq:ortho}
    T_{\text{ortho}}(k)=\sum_i c_ib^\prime_i.
\end{equation}
The validity of this transformation is proven in \ref{subsec: appendix ortho proof}.
\subsection{Some discussions}\label{subsec: discussion 3}
The above two gauge choices may have different results but both satisfy \eqref{eq:kinematic_constraint}. Usually, one can freely choose one of them. There are rare cases that one of the gauge choices become unavailable, when the corresponding local Gram matrix has a vanishing determinant. In practice, for the delta plane projection gauge, this happens when $n=1$ and $\Delta u_1$ is light-like. For the orthogonalization gauge, this happens for $n=1$ and $u_1$, thus also, $u_1^\prime$, is light-like. If one of the gauges is unavailable, one can switch to the other one. 


Besides, for some cases, the symmetry relations between individual integrals exist, but the symmetry relations between corresponding sectors cannot be found using the above methods. For example, for the triangle-bubble diagram in Fig. \ref{subfig: 2. tri bub}, consider its two sub sectors, a sunset diagram with the massless external legs in Fig. \ref{subfig: 2. sunset 1}, and a vacuum sunset diagram in Fig. \ref{subfig: 2. sunset 2}. Obviously, at least the corner integrals in the two sunset sectors are equal. However, the possible symmetry relation between the two sectors cannot be found with the method in this section. 
The corresponding momentum transformation which satisfies \eqref{eq: momentum symmetry trans Det. condition}, \eqref{eq:kinematic_constraint} and \eqref{eq:momentum symmetry trans prop. condition} does not exist.

\begin{figure}[hbtp]
\centering
\subfloat[triangle-bubble]{
\begin{tikzpicture}[scale=0.8]
\draw[line width=2pt] (0,0) arc (150:90:2);
\draw[line width=2pt] (0,0) arc (-90:-30:2);
\draw[line width=2pt](1.732,-1)--(0,0);
\draw[line width=2pt](1.732,-1)--(1.732,1.03);
\draw(1.732,-1)--(2.732,-1.5);

\draw(-1,-0.5)--(0,0);
\draw(-1,0.5)--(0,0);

\node[font=\large\bfseries] at (-1.5,-1){$1$};
\node[font=\large\bfseries] at (-1.5,1){$2$};
\node[font=\large\bfseries] at (3,-1.5){$3$};
\end{tikzpicture}
\label{subfig: 2. tri bub}
}
\hspace{5pt}
\subfloat[sunset 1]{
\begin{tikzpicture}[scale=0.8]
\draw[line width=2pt] (0,0) arc (120:60:2);
\draw[line width=2pt] (0,0) arc (-120:-60:2);
\draw[line width=2pt](2,0)--(0,0);
\draw(2,0)--(3,0);
\draw(-1,-0.5)--(0,0);
\draw(-1,0.5)--(0,0);
\node[font=\large\bfseries] at (-1.5,-1){$1$};
\node[font=\large\bfseries] at (-1.5,1){$2$};
\node[font=\large\bfseries] at (3,-0.5){$3$};
\end{tikzpicture}
\label{subfig: 2. sunset 1}
}
\hspace{10pt}
\subfloat[sunset 2]{
\begin{tikzpicture}[scale=0.8]
\draw[line width=2pt] (0,0) arc (120:60:2);
\draw[line width=2pt] (0,0) arc (-120:-60:2);
\draw[line width=2pt](2,0)--(0,0);
\draw(-1,0)--(0,0);
\draw(-1,-0.5)--(0,0);
\draw(-1,0.5)--(0,0);
\node[font=\large\bfseries] at (-1.5,-1){$3$};
\node[font=\large\bfseries] at (-1.5,1){$2$};
\node[font=\large\bfseries] at (-1.5,0){$1$};
\end{tikzpicture}
\label{subfig: 2. sunset 2}
}
\caption{A massive triangle-bubble diagram and its two sub diagrams}

\end{figure}
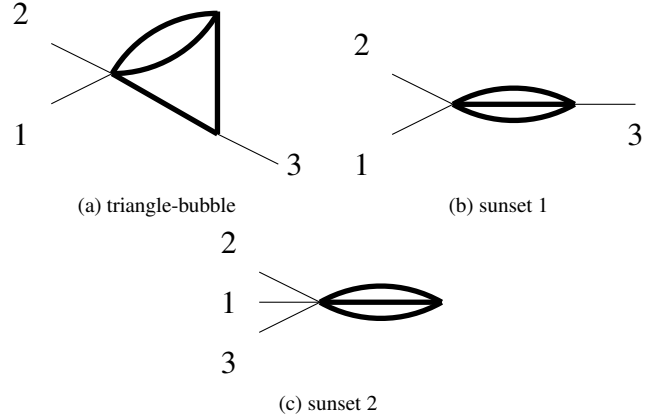\label{fig:A massive triangle-bubble diagram and its two sub diagrams}


Consequently, in this case the symmetry between corner integrals
cannot be lifted to the symmetry between two sectors. In package {\sc NeatIBP} \cite{Wu:2023upw}, for example, there is no effort to find the symmetry between these two sectors. Instead, the integral-level symmetry relations are obtained between the master integrals, by other approaches such as the Lee-Pomeransky representation \cite{Lee:2013hzt,Lee:2013mka,Lee:2014tja}.

\section{Determining external momentum groups}\label{sec:4}
We have introduced the method to find symmetry transformation in case some external momenta are grouped. 
For computational purposes, the momenta grouping information should be extracted from the propagator input, not the graphic representation of Feynman diagrams.  
In this section, we introduce two algorithms for this purpose.   
\subsection{Via momentum expression of propagators}

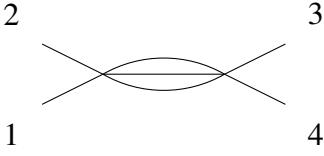
\begin{figure}[hbtp]
\centering

\begin{tikzpicture}[scale=0.8]
\draw (0,0) arc (120:60:2);
\draw(0,0) arc (-120:-60:2);
\draw(2,0)--(0,0);
\draw(2,0)--(3,0.5);
\draw(2,0)--(3,-0.5);

\draw(-1,-0.5)--(0,0);
\draw(-1,0.5)--(0,0);

\node[font=\large\bfseries] at (-1.5,-1){$1$};
\node[font=\large\bfseries] at (-1.5,1){$2$};
\node[font=\large\bfseries] at (3.5,1){$3$};
\node[font=\large\bfseries] at (3.5,-1){$4$};
\end{tikzpicture}

\caption{A massless sunset diagram}
\label{fig:4.1}
\end{figure}

We can determine the momentum grouping directly by the propagators. To do so, write down a matrix with the row index as a tuple $(i,j)$, and $k$ the column index,
\begin{equation}
    M_{(i,j),k}=\frac{\partial D_i}{\partial(l_j\cdot p_k)}
\end{equation}
For example, for a sunset diagram shown in Fig. \ref{fig:4.1} with propagators
\begin{equation}
    D_1=l_1^2,\,D_2=l_2^2,\,D_3=(l_1+l_2+p_1+p_2)^2,
\end{equation}

The matrix $M$ is
\begin{equation}
    M=\begin{pmatrix}
        0&0&0\\
        0&0&0\\
        0&0&0\\
        0&0&0\\
        2&2&0\\
        2&2&0
    \end{pmatrix}.
\end{equation}

After the above steps, we apply a row reduction on the above matrix, each nonzero row of the reduced matrix $(r_{(i,j)})_k$ corresponds to an external momentum group $\sum_k (r_{(i,j)})_k p_k$. In the above example, this gives only one group, as $\{p_1+p_2\}$. This means that this diagram depends only on $p_1+p_2$. 

However, sometimes this method may mistakenly consider some groups as separated. For example, in the diagram in Fig. \ref{fig:4.1}, the propagators can be redefined as
\begin{equation}
    D_1=(l_1-p_1)^2,\,D_2=(l_2-p_2)^2,\,D_3=(l_1+l_2)^2,
\end{equation}
The matrix $M$ is then
\begin{equation}
    M=\begin{pmatrix}
        -2&0&0\\
        0&0&0\\
        0&0&0\\
        0&-2&0\\
        0&0&0\\
        0&0&0
    \end{pmatrix},
\end{equation}
and the resulting groups are $\{p_1,p_2\}$. 
In order to fix this problem, we introduce additional requirements on the definition of propagators, such that for an $L$-loop diagram, there should be at least $L$ quadratic propagators, with momentum flow\footnote{We define the momentum flow for a quadratic propagator $k^2-m^2$ as $k$.} $l_1,l_2,\cdots,l_L$. If not, a transformation of the loop momenta would be applied to ensure this condition. We can always do this, unless the number of quadratic propagators with independent momentum flow is less than the number of loops. For these special cases, we discuss them in Section \ref{sec: discussion 4}. 
This algorithm is implemented in the package {\sc NeatIBP} \cite{Wu:2023upw} (including the earlier versions), and till now, we have not encountered any mistaken grouping.

\subsection{Via Feynman parameterization}

In this sub section, we introduce an algorithm to derive the momentum group by Feynman parameterization. This algorithm is newly implemented in {\sc NeatIBP} since version 1.0.4.7. 

Given a set of propagators, we can derive the Feynman parameterization and the Symanzik polynomials $\mathcal{U}$ and $\mathcal{F}$. Since the $p_i$-dependence of a Feynman integral is encoded in polynomial $\mathcal F$, we take the coefficients of each term in $\mathcal{F}$, stored as a list $\gamma_i$. As usual, these coefficients are linear in the Mandelstam variables and masses. Then we can derive a matrix $M$, with row index as a tuple $(i,j)$ and $k$ the column index, such that
\begin{equation}\label{eq: deriving-grouping matrix Feyn par}
    \frac{\partial \gamma_i}{\partial(p_k^\mu)}=\sum_j M_{(i,j),k} (p_j)_\mu.
\end{equation}

\begin{figure}[hbtp]
\centering

\begin{tikzpicture}[scale=0.8]

\draw(1.732,1)--(0,0);
\draw(1.732,-1)--(0,0);

\draw(1.732,-1)--(1.732,1);
\draw(1.732,1)--(2.732,1.5);
\draw(1.732,-1)--(2.732,-1.5);

\draw(-1,-0.5)--(0,0);
\draw(-1,0.5)--(0,0);

\node[font=\large\bfseries] at (-1.5,-1){$2$};
\node[font=\large\bfseries] at (-1.5,1){$3$};
\node[font=\large\bfseries] at (3,1.5){$4$};
\node[font=\large\bfseries] at (3,-1.5){$1$};

\end{tikzpicture}

\caption{A massless triangle diagram}
\label{fig:4.2}
\end{figure}
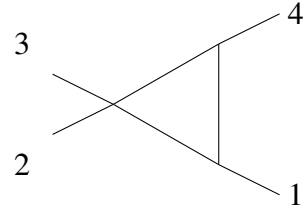

One very important point about the above step is that, when expressing $\mathcal F$ and taking the derivatives, we ignore all kinematic rules for external momenta. 
For example, for a massless triangle diagram with propagators shown in Fig. \ref{fig:4.2} with propagators
\begin{equation}
    D_1=l^2, \, D_2=(l+p_1)^2,\, D_3=(l+p_1+p_2+p_3)^2,
\end{equation}

We have
\begin{equation}
    {\mathcal F}=-p_1^2x_1x_2-(p_2+p_3)^2x_2x_3-(p_1+p_2+p_3)^2x_1x_3.
\end{equation}
since kinematic rules like $p_1^2=0$ or $p_4^2=0$ is ignored.
Thus, we can label 
\begin{equation}
    \gamma_1=-p_1^2,\, \gamma_2=-(p_2+p_3)^2,\,\gamma_3=-(p_1+p_2+p_3)^2.
\end{equation}
and
\begin{equation}
    \frac{\partial \gamma_1}{\partial(p_1^\mu)}=\frac{\partial (-p_1^2)}{\partial(p_1^\mu)}=2(p_1)_\mu.
\end{equation} 

So the matrix $M$ reads,
\begin{equation}
M=\begin{pmatrix}
        2&0&0\\
        0&0&0\\
        0&0&0\\
        0&0&0\\
        0&2&2\\
        0&2&2\\
        0&0&0\\
        2&2&2\\
        2&2&2\\
        2&2&2
    \end{pmatrix}
\end{equation}

After the above steps, we apply a row reduction on the matrix $M$, and each nonzero row of the reduced matrix $(r_{(i,j)})_k$ corresponds to an external momentum group $\sum_k (r_{(i,j)})_k p_k$. In the above example, this gives two momenta groups as $\{p_1,p_2+p_3\}$.
\subsection{Some discussions}\label{sec: discussion 4}

We need to state that, the above two methods are both applicable for integral families with linear propagators, as long as the number of independent\footnote{By independent, we mean the corresponding momenta flow are linearly independent.} quadratic propagators is not less than the number of loops $L$. If so, for the first method, we can always find enough quadratic propagators to be labeled as $l_1,l_2,\cdots,l_L$. For the cases where the number of independent quadratic propagators is less than the number of loops, the first method is not applicable. However, these cases are harmless 
because they corresponds to zero sectors.In these cases, we have the first Symanzik polynomial $\mathcal{U}= 0$. Thus, the Lee-Pomeransky representation \cite{Lee:2013hzt,Lee:2013mka,Lee:2014tja} polynomial\footnote{Although we cannot directly derive $\mathcal{F}$ from the propagator expression of the current sector since $\mathcal{U}=0$, we can derive it from their super sectors by setting some $x_i=0$.} $\mathcal{G}(x_i):=\mathcal{U}+\mathcal{F}=\mathcal{F}$. From the discussion in ref \cite{Lee:2008tj,Lee:2013mka}, if there is a set of $x$-independent $k_i$ such that
\begin{equation}
    \sum_{i} k_ix_i \partial_{x_i}\mathcal{G}=\mathcal{G},
\end{equation}
the corresponding sector is a zero sector. Since $\mathcal{G}=\mathcal{F}$ which is a degree-$(L+1)$ homogeneous polynomial, such $k_i$ always exist as $k_i=\frac{1}{L+1}$. Thus, the corresponding sectors are zero sectors.

\section{Example}\label{sec:example}
In this section, we present an example of this new method. Our example is from Ref. \cite{Badger:2024fgb}. This work reports that using {\sc NeatIBP} \cite{Wu:2023upw}, which employs the new symmetry methods,  more symmetry relations are found. This decreases the number of master integrals by $2$ compared to that from {\sc LiteRed} \cite{Lee:2008tj,Lee:2012cn,Lee:2013mka}, for two of the two-loop five-point integral families. In this section, we select one of the families, and show the detail of the symmetries found by the new algorithm. 

\begin{figure}[hbtp]
\centering
\begin{tikzpicture}[scale=1]
\draw[line width=2pt] (1.4,-0.5) -- (0.59,-0.81);
\draw (0,0) -- (0.59,-0.81);
\draw(0,0) -- (0.59,0.81);
\draw(1.4,0.5) -- (0.59,0.81);
\draw (1.4,0.5) -- (2.4,0.5);
\draw(2.4,0.5) -- (2.4,-0.5);
\draw[line width=2pt] (1.4,-0.5) -- (2.4,-0.5);
\draw (1.4,0.5) -- (1.4,-0.5);

\draw[line width=2pt] (0.59,-1.61) -- (0.59,-0.81);
\draw(-1,0)--(0,0);
\draw (0.59,1.61) -- (0.59,0.81);
\draw(2.4,0.5)--(3.2,0.8);
\draw[line width=2pt](2.4,-0.5)--(3.2,-0.8);

\node[font=\large\bfseries] at (0.3,-1.7){$1$};
\node[font=\large\bfseries] at (-1.3,0){$2$};
\node[font=\large\bfseries] at (0.3,1.7){$3$};
\node[font=\large\bfseries] at (3.5,1){$4$};
\node[font=\large\bfseries] at (3.5,-1){$5$};
\end{tikzpicture}
\caption{A pentagon-box diagram}
\label{fig:pbb}
\end{figure}
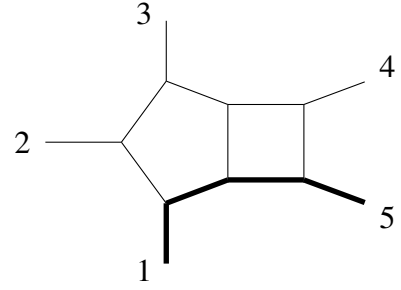

The diagram we are considering is the massive pentagon-box diagram shown in Fig. \ref{fig:pbb}. The propagators are defined as
\begin{equation}
    \begin{aligned}
        &D_1=l_1^2-m^2,\quad
        D_2=(l_1+p_1)^2,\quad
        D_3=(l_1+p_1+p_2)^2,\\
        &D_4=(l_1+p_1+p_2+p_3)^2,\quad
        D_5=(l_2-p_1-p_2-p_3)^2,\\
        &D_6=(l_2+p_5)^2,\quad
        D_7=l_2^2-m^2,\quad
        D_8=(l_1+l_2)^2,\\
        &D_9=(l_2+p_1)^2,\quad
        D_{10}=(l_2+p_1+p_2)^2,\\
        &D_{11}=(l_1+p_5)^2.
    \end{aligned}
\end{equation}
Among them, from $D_1$ to $D_8$ are propagators, and the rest three are irreducible scalar products. The external kinematics conditions are
\begin{equation}
    \begin{aligned}
        &p_1^2=p_5^2=m^2,\quad
        p_2^2=p_3^2=p_4^2=0,\\
        &(p_1+p_2)^2=s_{12},\quad
        (p_2+p_3)^2=s_{23},\quad
        (p_3+p_4)^2=s_{34},\\
        &(p_4+p_5)^2=s_{45},\quad
        (p_1+p_5)^2=s_{15},\quad
    \end{aligned}
\end{equation}
with the momentum conservation $p_1+p_2+p_3+p_4+p_5=0$.

We ran the reduction using {\sc NeatIBP}, we got $121$ master integrals. Then, we ran the reduction using {\sc LiteRed}+{\sc FIRE6} \cite{Smirnov:2019qkx}. In the latter framework, it uses {\sc LiteRed} to generate symmetry relations between sectors. The resulting number of master integrals is $123$. This agrees with what is stated in Ref. \cite{Badger:2024fgb}. 

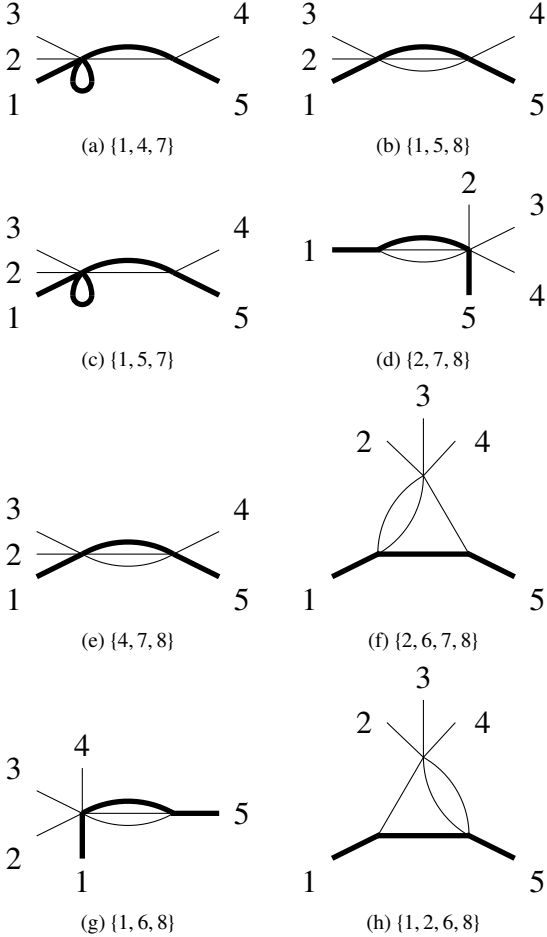
\begin{figure}[hbtp]
\centering
\subfloat[$\{1,4,7\}$]{

\begin{tikzpicture}[scale=0.6]
\draw[line width=2pt] (0,0) arc (120:60:2);
\draw[line width=2pt] (0,0) arc (135:180:0.707);
\draw[line width=2pt] (0,0) arc (45:0:0.707);
\draw[line width=2pt] (-0.207,-0.49) arc (-180:0:0.207);
\draw(2,0)--(0,0);

\draw(2,0)--(3,0.5);
\draw[line width=2pt](2,0)--(3,-0.5);
\draw[line width=2pt](-1,-0.5)--(0,0);
\draw(-1,0.5)--(0,0);
\draw(-1,0)--(0,0);
\node[font=\large\bfseries] at (-1.5,-1){$1$};
\node[font=\large\bfseries] at (-1.5,1){$3$};
\node[font=\large\bfseries] at (-1.5,0){$2$};
\node[font=\large\bfseries] at (3.5,1){$4$};
\node[font=\large\bfseries] at (3.5,-1){$5$};
\end{tikzpicture}
\label{subfig:pbb147}
}
\hspace{5pt}
\subfloat[$\{1,5,8\}$]{

\begin{tikzpicture}[scale=0.6]
\draw [line width=2pt](0,0) arc (120:60:2);
\draw (0,0) arc (-120:-60:2);
\draw(2,0)--(0,0);

\draw(2,0)--(3,0.5);
\draw[line width=2pt](2,0)--(3,-0.5);
\draw[line width=2pt](-1,-0.5)--(0,0);
\draw(-1,0.5)--(0,0);
\draw(-1,0)--(0,0);
\node[font=\large\bfseries] at (-1.5,-1){$1$};
\node[font=\large\bfseries] at (-1.5,1){$3$};
\node[font=\large\bfseries] at (-1.5,0){$2$};
\node[font=\large\bfseries] at (3.5,1){$4$};
\node[font=\large\bfseries] at (3.5,-1){$5$};

\end{tikzpicture}
\label{subfig: 2. pbb158}
}
\hspace{5pt}
\subfloat[$\{1,5,7\}$]{
\begin{tikzpicture}[scale=0.6]
\draw[line width=2pt] (0,0) arc (120:60:2);
\draw[line width=2pt] (0,0) arc (135:180:0.707);
\draw[line width=2pt] (0,0) arc (45:0:0.707);
\draw[line width=2pt] (-0.207,-0.49) arc (-180:0:0.207);
\draw(2,0)--(0,0);

\draw(2,0)--(3,0.5);
\draw[line width=2pt](2,0)--(3,-0.5);
\draw[line width=2pt](-1,-0.5)--(0,0);
\draw(-1,0.5)--(0,0);
\draw(-1,0)--(0,0);
\node[font=\large\bfseries] at (-1.5,-1){$1$};
\node[font=\large\bfseries] at (-1.5,1){$3$};
\node[font=\large\bfseries] at (-1.5,0){$2$};
\node[font=\large\bfseries] at (3.5,1){$4$};
\node[font=\large\bfseries] at (3.5,-1){$5$};
\end{tikzpicture}
\label{subfig:pbb157}
}
\hspace{5pt}
\subfloat[$\{2,7,8\}$]{

\begin{tikzpicture}[scale=0.6]
\draw [line width=2pt](0,0) arc (120:60:2);
\draw (0,0) arc (-120:-60:2);
\draw(2,0)--(0,0);

\draw(2,0)--(3,-0.5);
\draw[line width=2pt](2,0.025)--(2,-1);
\draw(2,0)--(3,0.5);
\draw(2,0)--(2,1);
\draw[line width=2pt](-1,0)--(0.06,0);

\node[font=\large\bfseries] at (-1.5,0){$1$};
\node[font=\large\bfseries] at (3.5,1){$3$};
\node[font=\large\bfseries] at (3.5,-1){$4$};
\node[font=\large\bfseries] at (2,1.5){$2$};
\node[font=\large\bfseries] at (2,-1.5){$5$};

\end{tikzpicture}
\label{subfig: 2. pbb278}
}
\hspace{5pt}
\subfloat[$\{4,7,8\}$]{
\begin{tikzpicture}[scale=0.6]
\draw [line width=2pt](0,0) arc (120:60:2);
\draw (0,0) arc (-120:-60:2);
\draw(2,0)--(0,0);
\draw(2,0)--(3,0.5);
\draw[line width=2pt](2,0)--(3,-0.5);
\draw[line width=2pt](-1,-0.5)--(0,0);
\draw(-1,0.5)--(0,0);
\draw(-1,0)--(0,0);
\node[font=\large\bfseries] at (-1.5,-1){$1$};
\node[font=\large\bfseries] at (-1.5,1){$3$};
\node[font=\large\bfseries] at (-1.5,0){$2$};
\node[font=\large\bfseries] at (3.5,1){$4$};
\node[font=\large\bfseries] at (3.5,-1){$5$};
\end{tikzpicture}
\label{subfig: 2. pbb478}
}
\hspace{5pt}
\subfloat[$\{2,6,7,8\}$]{
\begin{tikzpicture}[scale=0.6]
\draw (0,0) arc (180:120:2);
\draw (0,0) arc (-60:0:2);
\draw (2,0) -- (1,1.732);

\draw[line width=2pt](0,0)--(2,0);

\draw (1,3) -- (1,1.732);
\draw (1.7,2.5) -- (1,1.732);
\draw (0.2,2.5) -- (1,1.732);
\draw[line width=2pt](1.95,0.025)--(3,-0.5);
\draw[line width=2pt](-1,-0.5)--(0.05,0.025);

\node[font=\large\bfseries] at (-1.5,-1){$1$};
\node[font=\large\bfseries] at (1,3.5){$3$};
\node[font=\large\bfseries] at (-0.3,2.5){$2$};
\node[font=\large\bfseries] at (2.3,2.5){$4$};
\node[font=\large\bfseries] at (3.5,-1){$5$};
\end{tikzpicture}
\label{subfig: 2. pbb2678}
}
\hspace{5pt}
\subfloat[$\{1,6,8\}$]{

\begin{tikzpicture}[scale=0.6]
\draw [line width=2pt](0,0) arc (120:60:2);
\draw (0,0) arc (-120:-60:2);
\draw(2,0)--(0,0);

\draw(0,0)--(-1,-0.5);
\draw[line width=2pt](0,0.025)--(0,-1);
\draw(0,0)--(-1,0.5);
\draw(0,0)--(0,1);
\draw[line width=2pt](3,0)--(1.94,0);

\node[font=\large\bfseries] at (3.5,0){$5$};
\node[font=\large\bfseries] at (-1.5,1){$3$};
\node[font=\large\bfseries] at (-1.5,-1){$2$};
\node[font=\large\bfseries] at (0,1.5){$4$};
\node[font=\large\bfseries] at (0,-1.5){$1$};

\end{tikzpicture}
\label{subfig: 2. pbb168}
}
\hspace{5pt}
\subfloat[$\{1,2,6,8\}$]{
\begin{tikzpicture}[scale=0.6]
\draw (2,0) arc (0:60:2);
\draw (2,0) arc (240:180:2);
\draw (0,0) -- (1,1.732);

\draw[line width=2pt](0,0)--(2,0);

\draw (1,3) -- (1,1.732);
\draw (1.7,2.5) -- (1,1.732);
\draw (0.2,2.5) -- (1,1.732);
\draw[line width=2pt](1.95,0.025)--(3,-0.5);
\draw[line width=2pt](-1,-0.5)--(0.05,0.025);

\node[font=\large\bfseries] at (-1.5,-1){$1$};
\node[font=\large\bfseries] at (1,3.5){$3$};
\node[font=\large\bfseries] at (-0.3,2.5){$2$};
\node[font=\large\bfseries] at (2.3,2.5){$4$};
\node[font=\large\bfseries] at (3.5,-1){$5$};
\end{tikzpicture}
\label{subfig: 2. pbb1268}
}

\caption{Some sub-sector diagrams of Fig. \ref{fig:pbb}}
\label{fig:pbb sub sectors}
\end{figure}

We analyzed how the master integrals are distributed in different sectors. The master integrals from  {\sc LiteRed}+{\sc FIRE6} result are in $68$ sectors. The master integrals from  {\sc NeatIBP} result are in $66$ sectors. Among them, there are $64$ sectors that appear in both results, with the same number of master integrals in each sector. The rest of the different sectors are shown in Table \ref{tab: ss diff}. The relevant diagrams are shown in Fig. \ref{fig:pbb sub sectors}.
\begin{table}[h]
    \centering
    \begin{tabular}{|c|c|c|}
    \hline
         sector & number of MIs&from which result\\
         \hline
    
         $\{1,4,7\}$ & $1$ & {\sc NeatIBP} \\
         \hline
         $\{1,5,8\}$ & $2$ & {\sc NeatIBP} \\
         \hline
 
         $\{1,5,7\}$ & $1$ & {\sc LiteRed}+{\sc FIRE6}\\
         \hline
         $\{2,7,8\}$ & $1$ & {\sc LiteRed}+{\sc FIRE6}\\
         \hline
         $\{4,7,8\}$ & $2$ & {\sc LiteRed}+{\sc FIRE6}\\
         \hline
         $\{2,6,7,8\}$ & $1$ & {\sc LiteRed}+{\sc FIRE6}\\
         \hline
         
    \end{tabular}
    \caption{The difference of the sector structure of master integrals.}
    \label{tab: ss diff}
\end{table}

The sectors $\{1,5,7\}$ and $\{4,7,8\}$ from {\sc LiteRed}+{\sc FIRE6} result are explainable without the new method. They can map to sectors $\{1,4,7\}$ and $\{1,5,8\}$ using symmetry transformations with coefficients as constants. So, the difference is just due to the different preference of independent sectors. The corresponding transformations are
\begin{equation}\label{eq:trans example 157}
    l_1\mapsto l_2,\quad
    l_2\mapsto -l_1,\quad
    p_i \mapsto p_i\text{ for }i\in\{1,2,3,5\},
\end{equation}
and
\begin{equation}\label{eq:trans example 478}
    l_1\mapsto -l_2,\quad
    l_2\mapsto -l_1,\quad
    p_i \mapsto p_i\text{ for }i\in\{1,2,3,5\},
\end{equation}
where \eqref{eq:trans example 157} maps sector $\{1,5,7\}$ to sector $\{1,4,7\}$ by 
\begin{equation}
    D_1\mapsto D_7,\quad
    D_5\mapsto D_4,\quad
    D_7\mapsto D_1,
\end{equation}
and \eqref{eq:trans example 478} maps sector $\{4,7,8\}$ to sector $\{1,5,7\}$ by 
\begin{equation}
    D_4\mapsto D_5,\quad
    D_7\mapsto D_1,\quad
    D_8\mapsto D_8.
\end{equation}

The effect of our new method appears in sectors $\{2,7,8\}$ and $\{2,6,7,8\}$. 
From our new method, the following transformations,
\begin{small}
\begin{equation}\label{eq: example rational function trans.}
\begin{aligned}
    &l_1 \mapsto l_2, \quad
    l_2 \mapsto l_1,  \quad
    p_1 \mapsto p_5,  \quad
    p_5\mapsto p_1,\\
    &p_2 \mapsto 
    p_2 +\frac{ 2 m^2 - 2 s_{12} - s_{15} + s_{34}}{4 m^2 - s_{15}}p_1 - 
  \frac{ 2 m^2 - 2 s_{12} - s_{15} + s_{34}}{4 m^2 - s_{15}}p_5,\\
 & p_3 \mapsto 
    p_3 + \frac{ 2 s_{12} + s_{23} - s_{34} - 2 s_{45}}{4 m^2 - s_{15}} p_1- 
  \frac{2 s_{12} + s_{23} - s_{34} - 2 s_{45}}{4 m^2 - s_{15}}p_5.
  \end{aligned}
\end{equation}
\end{small}
satisfies \eqref{eq:kinematic_constraint} and transforms the propagators as
\begin{equation}\label{eq:2678 to 6218}
    D_2\mapsto D_6,\quad
    D_6\mapsto D_2,\quad
    D_7\mapsto D_1,\quad
    D_8\mapsto D_8.
\end{equation}
Thus, this transformation maps sector $\{2,7,8\}$ to sector $\{1,6,8\}$, and maps sector $\{2,6,7,8\}$ to sector $\{1,2,6,8\}$. See Fig. \ref{fig:pbb sub sectors} for the corresponding diagrams. Note that the coefficients of the linear combinations in \eqref{eq: example rational function trans.} {\it do} have rational functions in Mandelstam variables and masses, and our method efficiently found such a transformation. This transformation eliminated two master integrals using the 
symmetry relations as
\begin{equation}\label{eq: master integral symmetry 1}
    I_{0,1,0,0,0,0,1,1,0,0,0}=I_{1, 0, 0, 0, 0, 1, 0, 1, 0, 0, 0},
\end{equation}
and
\begin{equation}\label{eq: master integral symmetry 2}
    I_{0,1,0,0,0,1,1,1,0,0,0}=I_{1, 1, 0, 0, 0, 1, 0, 1, 0, 0, 0}.
\end{equation}

Notice that, in the above, we did not use \texttt{FindRules} in {\sc FIRE6}. This is a function that finds symmetry relations between individual integrals, rather than between sectors, by comparing their parameterized representations. One can run it to find symmetry relations between the master integrals after the IBP reduction. In this example, \eqref{eq: master integral symmetry 1} and \eqref{eq: master integral symmetry 2} can be found by \textit{integral-level symmetry} searching by \texttt{FindRules}. However, as stated, this is performed after the whole IBP reduction is finished. Without finding the \textit{sector-level symmetries}, we spent double/multiple efforts during the IBP reduction in the sectors that are actually related by these symmetry relations. The additional efforts are considerable when facing very complicated integral families. 

In this example, five additional sector-level symmetry maps found by the new method. Thus, they decrease the number of unique sectors by five. Specifically, {\sc LiteRed} found 136 unique sectors and {\sc NeatIBP} found 131. We list the five sector-level symmetry maps in Tab. \ref{tab: more symmetries}. Among them, the maps $\{2,7,8\}\mapsto \{1,6,8\}$ and $\{2,6,7,8\}\mapsto \{1,2,6,8\}$ are discussed above. The rest three maps does not appear in Tab. \ref{tab: ss diff} because there is no master integral on them. See Fig. \ref{fig:pbb sub sectors 2} for the diagrams corresponding to the three rest maps.

\begin{table}[h]
    \centering
    \begin{tabular}{|c|c|}
    \hline
         mapped sector & maps to\\
         \hline
         \{1,6,7\} & \{1,2,7\} \\
         \hline
         \{2,7,8\} & \{1,6,8\} \\
         \hline
         \{1,6,7,8\} & \{1,2,7,8\} \\
         \hline
         \{2,6,7,8\} & \{1,2,6,8\} \\
         \hline
         \{3,5,6,8\} & \{3,4,6,8\} \\
         \hline
    \end{tabular}
    \caption{The symmetry maps between sectors. }
    \label{tab: more symmetries}
\end{table}

\begin{figure}[hbtp]
\centering
\subfloat[$\{1,6,7\}$]{
\begin{tikzpicture}[scale=0.6]
\draw[line width=2pt] (0,0) arc (120:60:2);
\draw[line width=2pt] (0,0) arc (135:180:0.707);
\draw[line width=2pt] (0,0) arc (45:0:0.707);
\draw[line width=2pt] (-0.207,-0.49) arc (-180:0:0.207);
\draw(2,0)--(0,0);
\draw[line width=2pt](1.976,0)--(3,0);
\draw[line width=2pt](-1,-0.6)--(0,0);
\draw(-1,0.6)--(0,0);
\draw(-1,0.2)--(0,0);
\draw(-1,-0.2)--(0,0);
\node[font=\large\bfseries] at (-1.5,-1){$1$};
\node[font=\large\bfseries] at (-1.5,-0.33){$2$};
\node[font=\large\bfseries] at (-1.5,0.33){$3$};
\node[font=\large\bfseries] at (-1.5,1){$4$};
\node[font=\large\bfseries] at (3.5,0){$5$};
\end{tikzpicture}
\label{subfig:pbb167}
}
\hspace{5pt}
\subfloat[$\{1,2,7\}$]{
\begin{tikzpicture}[scale=0.6]
\draw[line width=2pt] (0,0) arc (120:60:2);
\draw[line width=2pt] (2,0) arc (135:180:0.707);
\draw[line width=2pt] (2,0) arc (45:0:0.707);
\draw[line width=2pt] (2-0.207,-0.49) arc (-180:0:0.207);
\draw(2,0)--(0,0);
\draw[line width=2pt](0.024,0)--(-1,0);
\draw[line width=2pt](3,-0.6)--(2,0);
\draw(3,0.6)--(2,0);
\draw(3,0.2)--(2,0);
\draw(3,-0.2)--(2,0);
\node[font=\large\bfseries] at (3.5,-1){$5$};
\node[font=\large\bfseries] at (3.5,-0.33){$4$};
\node[font=\large\bfseries] at (3.5,0.33){$3$};
\node[font=\large\bfseries] at (3.5,1){$2$};
\node[font=\large\bfseries] at (-1.5,0){$1$};
\end{tikzpicture}
\label{subfig:pbb127}
}
\hspace{5pt}
\subfloat[$\{1,6,7,8\}$]{
\begin{tikzpicture}[scale=0.6]
\draw[line width=2pt] (0,0) arc (120:60:2);
\draw (0,0) arc (-120:-60:2);
\draw (0,0) arc (240:330:{sqrt(3) - 1});
\draw[line width=2pt](1.976,0)--(3,0);
\draw[line width=2pt](-1,-0.6)--(0,0);
\draw(-1,0.6)--(0,0);
\draw(-1,0.2)--(0,0);
\draw(-1,-0.2)--(0,0);
\node[font=\large\bfseries] at (-1.5,-1){$1$};
\node[font=\large\bfseries] at (-1.5,-0.33){$2$};
\node[font=\large\bfseries] at (-1.5,0.33){$3$};
\node[font=\large\bfseries] at (-1.5,1){$4$};
\node[font=\large\bfseries] at (3.5,0){$5$};
\end{tikzpicture}
\label{subfig:pbb1678}
}
\hspace{5pt}
\subfloat[$\{1,2,7,8\}$]{
\begin{tikzpicture}[scale=0.6]
\draw[line width=2pt] (0,0) arc (120:60:2);
\draw (0,0) arc (-120:-60:2);
\draw (2,0) arc (-60:-150:{sqrt(3) - 1});
\draw[line width=2pt](0.024,0)--(-1,0);
\draw[line width=2pt](3,-0.6)--(2,0);
\draw(3,0.6)--(2,0);
\draw(3,0.2)--(2,0);
\draw(3,-0.2)--(2,0);
\node[font=\large\bfseries] at (3.5,-1){$5$};
\node[font=\large\bfseries] at (3.5,-0.33){$4$};
\node[font=\large\bfseries] at (3.5,0.33){$3$};
\node[font=\large\bfseries] at (3.5,1){$2$};
\node[font=\large\bfseries] at (-1.5,0){$1$};
\end{tikzpicture}
\label{subfig:pbb1278}
}
\hspace{5pt}

\subfloat[$\{3,5,6,8\}$]{
\begin{tikzpicture}[scale=0.6]
\draw (0,0) arc (150:90:2);
\draw (0,0) arc (-90:-30:2);
\draw (0,0) -- (1.732,-1) -- (1.732,1);
\draw (1.732-0.02,1-0.02*1.732) -- (1.732+0.5,1+0.5*1.732);
\draw (1.732-0.02,-1+0.02*1.732) -- (1.732+0.5,-1-0.5*1.732);
\draw(-1,0.5)--(0,0);
\draw[line width=2pt](-1,0)--(0,0);
\draw[line width=2pt](-1,-0.5)--(0,0);
\node[font=\large\bfseries] at (-1.5,-1){$5$};
\node[font=\large\bfseries] at (-1.5,0){$1$};
\node[font=\large\bfseries] at (-1.5,1){$2$};
\node[font=\large\bfseries] at (1.732+1,1+0.5*1.732){$3$};
\node[font=\large\bfseries] at (1.732+1,-1-0.5*1.732){$4$};
\end{tikzpicture}
\label{subfig:pbb3568}
}
\hspace{5pt}
\subfloat[$\{3,4,6,8\}$]{
\begin{tikzpicture}[scale=0.6]
\draw (0,0) arc (-150:-90:2);
\draw (0,0) arc (90:30:2);
\draw (0,0) -- (1.732,1) -- (1.732,-1);
\draw (1.732-0.02,1-0.02*1.732) -- (1.732+0.5,1+0.5*1.732);
\draw (1.732-0.02,-1+0.02*1.732) -- (1.732+0.5,-1-0.5*1.732);
\draw[line width=2pt](-1,0.5)--(0,0);
\draw[line width=2pt](-1,0)--(0,0);
\draw(-1,-0.5)--(0,0);
\node[font=\large\bfseries] at (-1.5,-1){$5$};
\node[font=\large\bfseries] at (-1.5,0){$1$};
\node[font=\large\bfseries] at (-1.5,1){$2$};
\node[font=\large\bfseries] at (1.732+1,1+0.5*1.732){$4$};
\node[font=\large\bfseries] at (1.732+1,-1-0.5*1.732){$3$};
\end{tikzpicture}
\label{subfig:pbb3468}
}
\hspace{5pt}
\caption{Some other sub-sector diagrams of Fig. \ref{fig:pbb}}
\label{fig:pbb sub sectors 2}
\end{figure}

We have provided a {\sc Mathematica} readable file including all the symmetries found by the new algorithm (implemented in {\sc NeatIBP}) for this example. The file name is \texttt{symmetries\_in\_the\_example.txt}. The data is a list of all possible propagator permutations that leads to sector-level symmetries. For each member in the list, the first entry is a rule of the permutation. For example, the permutation in \eqref{eq:2678 to 6218} is recorded as $\{2,6,7,8\}\to\{6,2,1,8\}$. Notice that there could be more than one permutation between two sectors. The second entry is the momentum transformation rule. The meaning of the notations in the file are shown in Tab. \ref{tab: notations}.

\begin{table}[h]
    \centering
    \begin{tabular}{|c|c|}
    \hline
         notations in the file & meaning\\
         \hline
         
         s12, ... , s15 &$s_{12},\cdots,s_{15}$\\\hline
         mm & $m^2$\\\hline
         
         l1,l2 & $l_1,l_2$\\\hline
         k1, ... , k5 & $p_1,\cdots, p_5$\\
         
         \hline
    \end{tabular}
    \caption{Notations in the file }
    \label{tab: notations}
\end{table}

\section{Summary}\label{sec:summary}

In this paper, we present a new method to derive more symmetry relations between Feynman integrals. They correspond to momentum transformation with coefficients as rational functions of kinematic parameters. These kinds of symmetry relations usually appear between low sub sectors of an integral family. In these cases, the external momenta are in groups. This introduces the continuous degree of freedom. By introducing certain gauge choices, the new method fixes the degree of freedom and produces the solution for the transformation efficiently.

The new method introduced in this paper leads to a systematic symmetry searching algorithm for an integral family. Compared to some state-of-the-art reduction programs, it finds more sector-level symmetry relations. Usually, this new algorithm makes itself useful when the considered sectors are very low sub sectors of a family, such that they have multiple external momenta injecting to a same vertex in the diagram, forming so-called momentum groups.

We remind that, although we used the implementation in {\sc NeatIBP} \cite{Wu:2023upw} to display the example in Section \ref{sec:example}, this algorithm can be implemented individually in one's own codes. We give a brief description of the procedure here:
\begin{enumerate}
    \item Determine the external momentum groups for the sectors being considered, using the algorithm introduced in Section \ref{sec:4}.
    \item Treat each momentum group as if it is a single external line, and treat the corresponding sub diagram as an $n_G$-point diagram, where $n_G$ is the number of groups. Use traditional symmetry algorithms to solve symmetry problems for the $n_G$-point diagrams. Transformations of the loop momenta and external momentum groups will be obtained, in the form shown in \eqref{eq:loop momentum symmetry trans. with momentum group} and \eqref{eq:external momentum symmetry trans. with momentum group}.
    \item Take the solution of external momentum group transformation, in the form shown in \eqref{eq:external momentum symmetry trans. with momentum group}, as an input. Use algorithms introduced in Section \ref{sec:3} to derive the transformation rule of the original external momenta. During this step, one can pick up one preferred method among the two introduced in Sectiorn \ref{sec:3}. If one encounters the Gram matrix with vanishing determinant, switch to another method.
    
\end{enumerate}

Using the above algorithms, we find more sector-level symmetry relations and decrease the number of sectors to be concerned in IBP reductions. 
Consequently, this new method boosts the IBP reduction process. Additionally, the idea of this new method also works for finding the equivalent sub sectors of different Feynman integral families.

\section*{Acknowledgement} 
We thank Fabian Lange, Roman Lee, Yan-Qing Ma, Johann Usovitsch and Simone Zoia for important discussions.
ZW is supported by The Hangzhou Human Resources and Social Security Bureau through The First Batch of Hangzhou Postdoctoral Research Funding in 2024.
YZ is supported by the NSF of China through
Grant No. 12047502, 12247103, and 12075234.

\appendix
\section{Several proofs}
\subsection{Proofs about the delta plane projection gauge}\label{subsec: appendix dpp proof}
Now we prove the method based on the delta plane projection gauge introduced in Section \ref{subsec:dpp}. There are two conditions to be checked:
\begin{enumerate}
    \item  $T_{\text {dpp}}(u_i)=u_i^\prime$, for $i\leq n$, and $i>n$.
    \item The transformation \eqref{eq:delta plane projection transformation} indeed satisfies \eqref{eq:kinematic_constraint}.
\end{enumerate}

We first check the first condition. For $i\leq n$, from the above discussions, together with $u_i^{\perp}\cdot \omega_j=0$, we have 
 \begin{equation}
 \begin{aligned}
     \sum_{j,k=1}^n (u_i\cdot \omega^j) {B_j}^k \omega_k=\sum_{j,k=1}^n{A_i}^j{B_j}^k \omega_k=\sum_{k=1}^n{A^\prime_i}^k \omega_k={u_i^\prime}^\parallel
\end{aligned}
 \end{equation}
 Since $(\Delta u_i)^\perp=0$, we have $u_i^\perp={u_i^\prime}^\perp$. Thus, $T_{\text {dpp}}(u_i)=u_i^\prime$, for $i\leq n$.
 
To check the rest part of the above conditions, we need to remember that
\begin{equation}\label{eq:up unchange-distance condition}
    u_i\cdot u_j=u_i^\prime \cdot u_j^\prime,
\end{equation}
for a valid transformation.
Thus, for $i,j\leq n$, together with some of the properties we have discussed above, we have
\begin{equation}
    \sum_{k,l=1}^n  {A_i}^k(\omega_k\cdot \omega_l){A_j}^l=\sum_{k,l=1}^n  {A_i^\prime}^k(\omega_k\cdot \omega_l){A_j^\prime}^l,
\end{equation}
Multiplying the both sides by the inverse of matrix $A$, we have
\begin{equation}\label{eq:B condition}
    \omega_i\cdot \omega_j=\sum_{k,l=1}^n  {B_i}^k(\omega_k\cdot \omega_l){B_j}^l.
\end{equation}
This is equivalent to 
\begin{equation}\label{eq:B inv}
    {(B^{-1})_i}^j={B^j}_i
\end{equation}

For $i\leq n$ and $a>n$, \eqref{eq:up unchange-distance condition} is 
\begin{equation}
     \sum_{k=1}^n  {A_i}^k\omega_k \cdot u_a^\parallel=\sum_{k=1}^n  {A_i^\prime}^k\omega_k \cdot{u_a^\prime}^\parallel.
\end{equation}
Multiplying the both sides by the inverse of $A$ matrix, we have
\begin{equation}\label{eq:ua transform}
   \omega_i \cdot u_a^\parallel=\sum_{j,k=1}^n  {B_i}^k\omega_k \cdot \omega_j({u_a^\prime}\cdot \omega^j)
\end{equation}
Now we can further check the two conditions. For the first condition while considering $T_{\text {dpp}}(u_a)$ for $a>n$, from \eqref{eq:B inv} and \eqref{eq:ua transform}, we have
\begin{equation}
\begin{aligned}
     &\sum_{i,j=1}^n (u_a\cdot \omega^i) {B_i}^j \omega_j=\sum_{i,j=1}^n (u_a^\parallel\cdot \omega_{i}) {B}^{ij} \omega_j\\&
     =\sum_{i,j,k,l=1}^n  {B_i}^{l}\omega_l \cdot \omega_k({u_a^\prime}\cdot \omega^k){B^i}_{j} \omega^j\\&
     =\sum_{k=1}^n ({u_a^\prime}\cdot \omega^k) \omega_k={u_a^\prime}^\parallel.
\end{aligned}
\end{equation}
Thus, we have $T_{\text{dpp}}(u_a)=u_a^\prime$, and the first condition is checked.

 We then check the second condition. For any vector $k$, we have
\begin{equation}
\begin{aligned}
T_{\text {dpp}}(k)^2&=(k^\perp)^2+\sum_{i,j,i^\prime,j^\prime=1}^n (k\cdot \omega^i){B_i}^j (\omega_j\cdot\omega_{j^\prime})(k\cdot \omega^{i^\prime}){B_{i^\prime}}^{j^\prime} \\
&=(k^\perp)^2+\sum_{i,i^\prime=1}^n (k\cdot \omega^i) (\omega_i\cdot\omega_{i^\prime})(k\cdot \omega^{i^\prime})\\
&=(k^\perp)^2+(k^\parallel)^2=k^2.
\end{aligned}
\end{equation}
Thus, the second condition is checked. Here we have used \eqref{eq:B condition}.
\subsection{Proofs about the orthogonalization gauge}\label{subsec: appendix ortho proof}
We now prove the method based on orthogonalization gauge introduced in Section \ref{subsec:ortho}. This is to check the two conditions:
\begin{enumerate}
    \item $T_{\text{ortho}}(u_i)=u_i^\prime$.
    \item $(T_{\text{ortho}}(k))^2=k^2$, for $\forall k$.
\end{enumerate}

For the first condition when $i\leq n$, the proof is trivial. When $a>n$, let
\begin{equation}
    u_a=\sum_{j=1}^n{c_a}^j u_j,
\end{equation}
and
\begin{equation}
    u_a^\prime=\sum_{j=1}^n{c^\prime_a}^j u^\prime_j,
\end{equation}
Considering \eqref{eq:up unchange-distance condition}, we have
\begin{equation}
  \sum_{j=1}^n{c_a}^j u_i\cdot u_j=\sum_{j=1}^n{c^\prime_a}^j u^\prime_i\cdot u^\prime_j.
\end{equation}
Multiplying the both sides by the inverse local gram matrix, which is $(u\cdot u)^{-1}=(u^\prime\cdot u^\prime)^{-1}$, we have
\begin{equation}
    {c^\prime_a}^i={c_a}^i.
\end{equation}
Thus, we have $T_{\text{ortho}}(u_a)=u_a^\prime$.

For the second condition, the proof is also obvious by using \eqref{eq:up unchange-distance condition}.

\section*{Declaration of generative AI and AI-assisted technologies in the writing process} 
During the preparation of this work the authors used Writefull Premium in order to perform spelling/grammar checks, and to improve language as well as readability of this paper. After using this service, the authors reviewed and edited the content as needed and takes full responsibility for the content of the publication.

\bibliographystyle{elsarticle-num}
\bibliography{bibtex}

\end{document}